# Programmable Charge Trap for Junction-less selective extraction of holes in Solar Cells

Swasti Bhatia, *Dept. of Electrical Engineering, Indian Institute of Technology Bombay, Mumbai India*,
Aldrin Antony, *Physics Department Cochin University of Science and Technology, Kochi, India*, and
Pradeep R. Nair, *Dept. of Electrical Engineering, Indian Institute of Technology Bombay, Mumbai*

*Abstract*: Selective extraction of photo-generated carriers is a fundamental challenge in solar cells which is usually achieved through junctions with the associated doping as well as band offset differences. In this context, here we propose a new paradigm for selective extraction for majority carriers through novel usage of the **programmable charge trap which comprises of Oxide-Nitride-Oxide (ONO) stack and has the primary function of holding electrically injected charge. Through detailed numerical simulations, here we show that such a charge trap with an additional metal contact can (i) compensate for efficiency loss due to sub-optimal passivation and sub-optimal hole selectivity in homojunction as well as transition metal oxide-based heterojunction solar cells and (ii) can also function as a standalone hole selection scheme. The proposed scheme, with its easy integration and the capability of programmable compensation of performance loss, is of interest to the photovoltaic community.**

## I. Introduction

Solar photovoltaics has shown great potential as reliable renewable energy source in the last two decades. The two key functions of any solar cell are to generate electron and hole pairs on absorbing solar spectra and then to collect the two polarities of charge at different metal contacts. For well-established and commercially available Si solar cells structures like PERC (Partial Emitter with Rear Contact) [1] and Aluminum BSF (Back Surface Field) [2][3], carrier separation is achieved from doped homojunctions. Such carrier extraction relies on difference in electrochemical potential on the two ends of the homojunction. However, in the last two decades, new technologies like heterojunction with intrinsic thin film (HIT) [4][5] solar cells and transition metal oxide (TMO) based heterojunction solar cells [6] have emerged which use a combination of asymmetric band offsets [7][8] and difference of electrochemical potential [9] to drive charge separation with excellent success [10]. Here, we introduce yet another selection mechanism where the selectivity comes from electric field created by intentional and programmable charge stored in an Oxide-Nitride-Oxide (ONO) stack (fig. 1). The key advantages of the proposed mechanism, as will be demonstrated in this work, are: its capability to recover performance loss via programming after deployment in the field and its capability to act as a standalone hole selection scheme – i.e., in the absence of any doped contacts irrespective of holes being majority or minority carriers.

An interface or bulk charge has been well known to provide field effect passivation[1][2] at heterojunctions of Si with oxides as in the case of $Al_2O_3$[3] and has also been explored for achieving charge selectivity[4]. While highly useful for the former application, the latter has had limited success even when the same is known to be suitable for low cost large area solar cells[5]. The primary problems that impede the realization of such devices is the high fixed densities of charge required to achieve selectivity and the degradation of that charge under solar radiation which also contains ultraviolet (UV) radiation[6]. The structure proposed here, has the potential to circumvent both the above-mentioned problems by using charge trap, the blocking oxides on either of which make the loss of programmed charge difficult instead of a single charged layer as proposed previously. In addition, there is also a charging contact deposited over the stack (fig. 1) that provides the capability to inject charge into the trap and hence the charge density available is no longer restricted by the amount of fixed charge that can be introduced while deposition. Further, the loss of trapped charge post deployment can be recovered through a re-charging or re-programming of the charge trap by applying a high enough voltage to the charging contact. The UV induced charge leakage is avoided by the integration of the structure at the rear end which is does not receive significant short wavelength radiation.

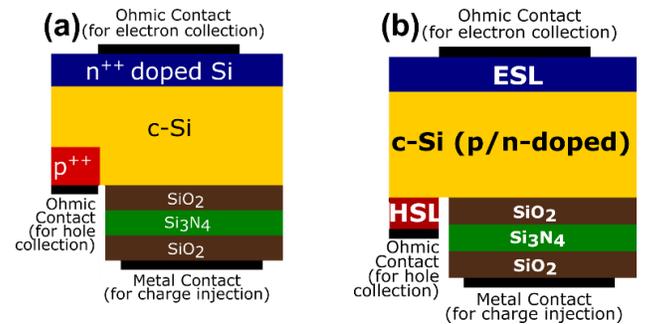

Fig. 1: Proposed structures of partially contacted charge trap based solar cells (not drawn to scale). The efficacy of the charge traps is explored for both (a) homojunction solar cells and (b) heterojunction solar cells based on silicon. ESL stands for Electron Selective Layer and HSL for hole selective layer. Thickness of films are not to scale.

Below, we first quantify the efficiency improvement brought about by the integration of the proposed scheme in transition metal oxide-based heterojunction (alternatively known as carrier selective contact or CSC solar cells) and PERC solar cells. Second, we demonstrate the effectiveness of the scheme to compensate for loss of performance post deployment. Third, and most importantly we test the performance limits of the stack if used as an independent hole selection scheme. Finally, we optimize the area fraction needed for best performance of the scheme.

## II. MODEL SYSTEM

**System Description:** The device structures shown in fig. 1 are modeled using Sentaurus TCAD[7]. The coupled electron and hole continuity and Poisson equations are solved self-consistently under illuminated and dark conditions. The illumination conditions are modeled by uniform generation throughout c-Si, which corresponds to 1-sun. Radiative and trap assisted Shockley-Read-Hall recombination mechanisms are considered inside Si bulk. The key material parameters used are listed in Appendix A. We test the effectiveness of an ONO stack with different negative charge densities for two different architectures of PERC and Si/TMO heterojunction solar cells.

**Working Principle:** The structures of a homojunction and heterojunction solar cell with a charge trap are shown in fig 1 (a) and 1(b) respectively. The band diagrams of the charge trap introduced (in the structures shown in fig. 1) are shown in fig. 2. The band diagram, obtained at a high forward bias of 650 mV clearly shows that in the presence of a negative charge density of $10^{18}$ cm$^{-3}$ in the SiN$_X$ layer, a small electric field in Si bulk is retained which is not the case for an uncharged trap under same applied forward bias. This field is expected to provide field effect interface passivation by repelling electrons away from the interface. Hence, the scheme enhances the performance of the structures shown in fig. 1. Therefore, in view of this evident advantage of negatively charged trap, we will now discuss the process and feasibility of the injecting negative charge in the trap.

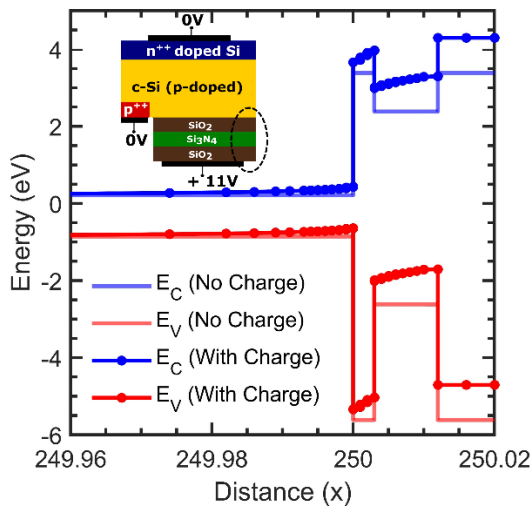

Fig. 2: Band diagram of a proposed solar cell at the rear side comparing the charged and uncharged trap at an applied forward bias of 650 mV. The negative charge density is kept constant for charged devices at $10^{19}$/cm$^3$. The depicted region corresponds to the region marked on the inset schematic on the top left.

The negative charge can be injected into the ONO trap through similar methods as employed in EEPROM memories – i.e., by applying a high enough positive voltage with respect to front electron contact. The charge injection can done effectively by utilizing the programming techniques developed for flash memories like the Incremental Step Pulse Programming (ISPP) [8][9]. The electrons reach SiN$_X$ from Si bulk via Fowler Nordheim (FN) tunneling. Fig. 3 shows the band diagram of the homojunction and the charge trap with a high bias of 11V applied to charging contact with the respect to the top electron contact. Note that even for such a high applied bias, the potential drop at the front homojunction is a negligible fraction of the total applied bias hence is not expected to retard the charge injection process.

**Charge Retention:** Once sufficient charge has been introduced into the charge trap as described above, the next challenge critical for the success of the scheme is the retention of this charge. The band diagram in the stack region near open circuit voltage (fig. 2) also shows that the bias does not bend the oxide bands enough to be able to cause any loss of charge. The injected charge is expected to have good retention as similar charge traps are known to have extensive use of such structure in flash memory devices[10]. Additionally, charge retention (quantified by $V_T$ (threshold voltage) change over time) and the efficiency of charge injection are improvised by the implementing advanced structures like MANOS[11] (Metal/AlOx/Nitride/Oxide/Silicon) and TANOS[12] (TaN/AlO$_X$/Nitride/Oxide/Silicon). Similar high retention structures are also relevant for improving charge retention in charge trap based solar cells proposed here. Furthermore, the issue of charge retention is expected to be significantly less critical for solar cells as compared to memories since the former does not undergo multiple program and erase cycles. However, some leakage of charge over a span of years is inevitable and the proposed scheme is designed to recover the same. The third charging contact in the proposed cell structure can also be used for re-injection of charge to compensate for leakage after deployment.

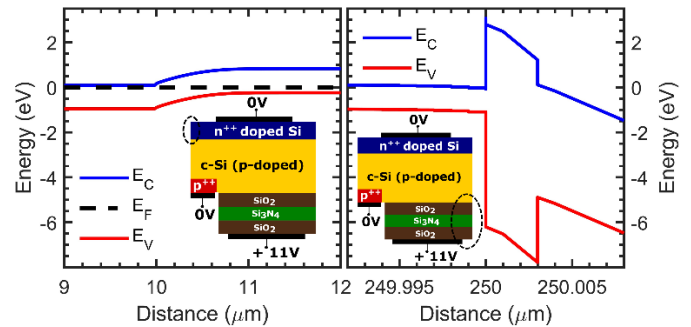

Fig. 3: Band diagram of the front homojunction and the rear charge trap with an applied bias of 11V at charging contact with respect to the front electron contact. The region of which the band diagram is plotted is shown in the inset schematic of each figure.

## III. RESULTS AND DISCUSSIONS

In this section we test the proposed charge-based hole selective solar cell structure for (i) its ability to improve the performance of existing heterojunction-based Si solar cells (ii) its potential to function as a standalone hole selectivity scheme. Finally, we discuss the role of area fraction which is a measure of the relative ratio of area contacted to area covered by ONO stack.

### A. Compensation of suboptimal passivation and hole selectivity

Si/TMO (transition metal oxide) based heterojunction solar cells have been explored extensively for their potential low fabrication costs and high efficiency. However, the TMOs of interest are not capable of providing a proper band offset for selection of one carrier along with optimal passivation. Hence, the best performing devices rely on an additional layer of a-Si for passivation [13][14]. The lack of passivation at the Si/HSL (Hole Selective Layer) has been modeled by increasing the surface recombination velocity (SRV) at the interface from 100 to 500 cm/s. Fig. 4 shows that the $V_{OC}$ (open circuit voltage) lost to interface recombination at Si/HSL can be fully compensated for by injection of negative charge of the order of $10^{18}$-$10^{19}$/cm$^3$ in the SiN$_X$ layer for SRV values as high as 500 cm/s. The results can be explained by the presence of an additional electric field, emanating from the programmed fixed negative charge in the dielectric stack. This electric field repels the electrons away from the Si/TMO interface and drives them deeper into the Si bulk. This deficiency of one type of carriers (electrons) at the Si interface, suppresses interface recombination even in the presence of interface traps. The charge in the trap therefore creates a field effect passivation at the Si/TMO. Interestingly, the field extends laterally also since the fixed charge is not directly beneath the Si/TMO interface. This field repels the electrons from the interface, and consequently the efficiency loss due the recombination is avoided even with a high SRV at the Si/HSL interface. These results illustrate the capability of the proposed structure to recover the efficiency loss due to interface degradation [15][16].

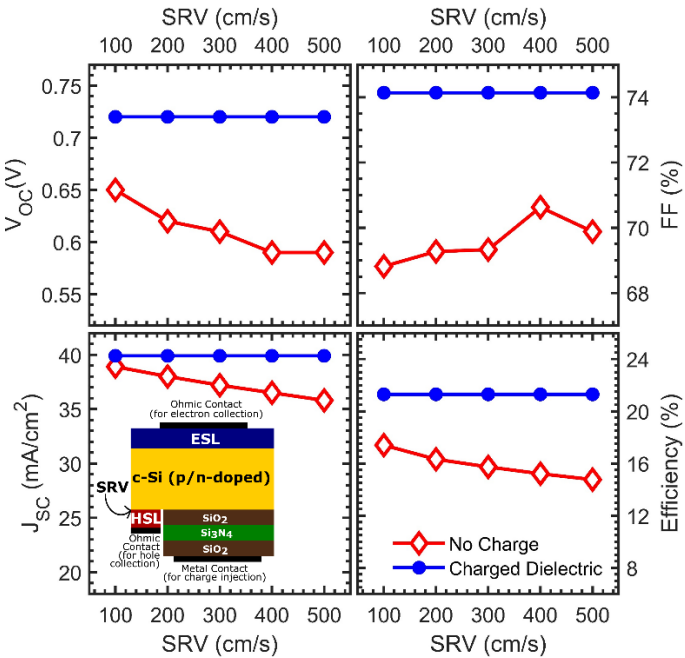

Fig. 4: Effect of high surface recombination velocity on solar cell performance parameters for charged and uncharged trap structures. The negative charge density is kept constant for charged devices at $10^{19}$/cm$^3$.

In addition to recombination, many of the TMOs used for hole selection like V$_2$O$_5$[17], MoO$_X$[18][19] and WO$_X$[20][21] do not have a large enough conduction band offset and rely only on electrochemical potential difference between Si and TMO to achieve hole selectivity. A detailed study on their selectivity has revealed that such a structure without a conduction band offset ($\Delta E_C$) to stop electrons is not capable of delivering $V_{OC}$ beyond 650 mV[22]. Further, none of the experimental devices using Si/TMO only for hole extraction have reached a $V_{OC}$ of 700 mV[23][24][18][25][26]. Hence, in this study we consider a HSL with no $\Delta E_C$ which covers 1% of the total rear side area. A comparison between a charged and an uncharged dielectric stack reported in fig. 3 clearly shows that with introducing a charge density of $10^{19}$/cm$^3$ in the stack, the performance loss due sub-optimal hole selectivity can be entirely compensated. Therefore, the scheme can be used with any passivating material regardless of $\Delta E_C$ to function as a perfect hole selective film independently. Hence, usage of the proposed charge based solar cells can significantly increase the number of options available for contact passivation.

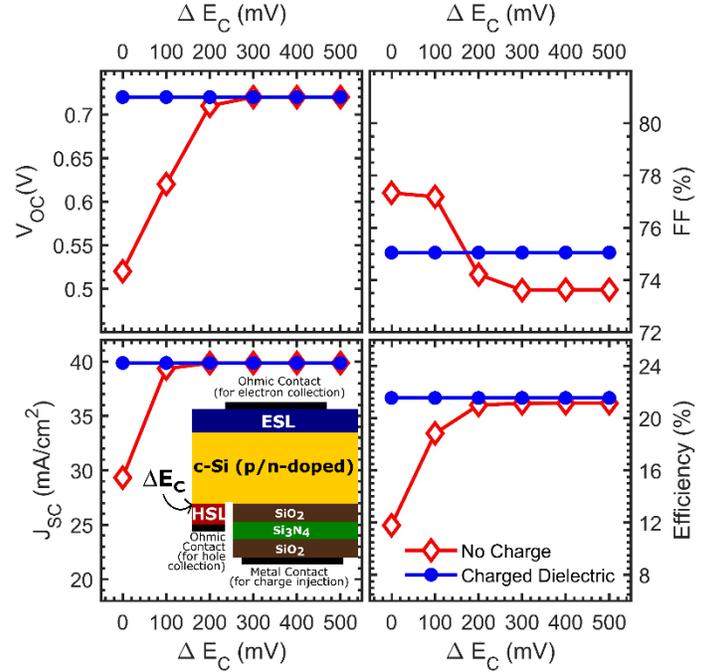

Fig. 5: Effect of varying conduction band offset on solar cell performance parameters for charge trap with and without negative charge. The negative charge density is kept constant for charged devices at $10^{19}$/cm$^3$.

### B. Charge Trap Based Hole Selectivity as an Independent Hole Selection Scheme

In the previous section we established the utility of the charged ONO stack when used in conjunction with Si/HSL (Hole Selective Layer) heterojunction for hole selection. The results of the previous section, especially the independence of performance of charge-based cells from $\Delta E_C$ as well as SRV, motivate an exploration of the proposed hole selectivity scheme as a standalone hole selective structure. Hence, we now study modified versions of structures shown in fig. 1(a) and fig. 1(b), i.e., with HSL layer and P+ doped silicon region respectively replaced with ohmic contact to c-Si bulk. We vary the negative charge density in the $SiN_X$ region and analyze the effect on solar cell performance for both p-doped and n-doped Si bulk.

We first investigate a solar cell with a Si/TMO heterojunction-based for electron extraction and charge-based scheme for hole extraction. The structure of the device is shown in fig. 6. These results show that for large charge densities (i.e., $10^{19}/cm^3$), the structure can indeed function as an effective hole selection scheme. We note that the charge densities equivalent to the values used in this work have been previously reported for $SiN_X$[27][28]. The charged stack serves as a minority carrier extraction scheme when the bulk Si is n-doped and as a majority carrier extraction scheme when the bulk Si is p-doped. The performance for a sufficiently charged stack is similar irrespective of whether it is being used as majority or minority charge selection mechanism for Si/TMO based heterojunction solar cells.

selection scheme can deliver high efficiency solar cell only if it is used as a minority carrier collecting scheme. The less-than-ideal performance of the trap-based structure as a majority carrier collection scheme can be attributed to the lower performance of the minority carrier extracting homojunction. The performance bottleneck comes from the doped homojunction rather than the trap-based selection mechanism. Hence, the scheme is useful for enhancing performance of an Al-BSF (Aluminum Back Surface Field) structure and is also capable of eliminating the rear P+ doped Si region. In conclusion, the results of Fig. 7 also suggest the charge-based hole selectivity scheme can be highly useful for fabrication process simplification of a n-Si based PERC (Passivated Emitter and Rear Contact) solar cells, since it has the potential to eliminate the localized p+ doping step.

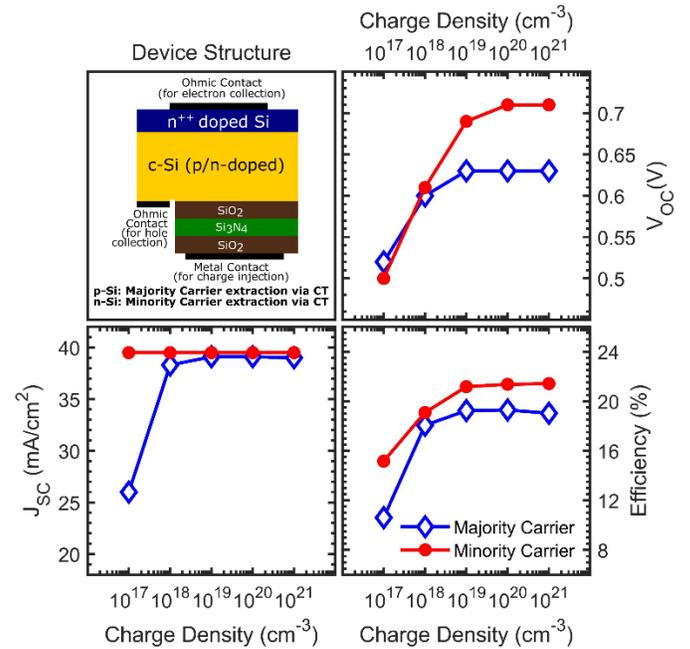

Fig. 7: Effect of charge density on device performance for a solar cell with n+-Si/p-Si and n+-Si/n-Si homojunction as an electron selection scheme and charges ONO stack as hole selective scheme. The charged stack is explored as a hole selection mechanism for a p-Si substrate when it functions as a majority carrier selection scheme and for a n-substrate Si where it functions as a minority carrier selection scheme.

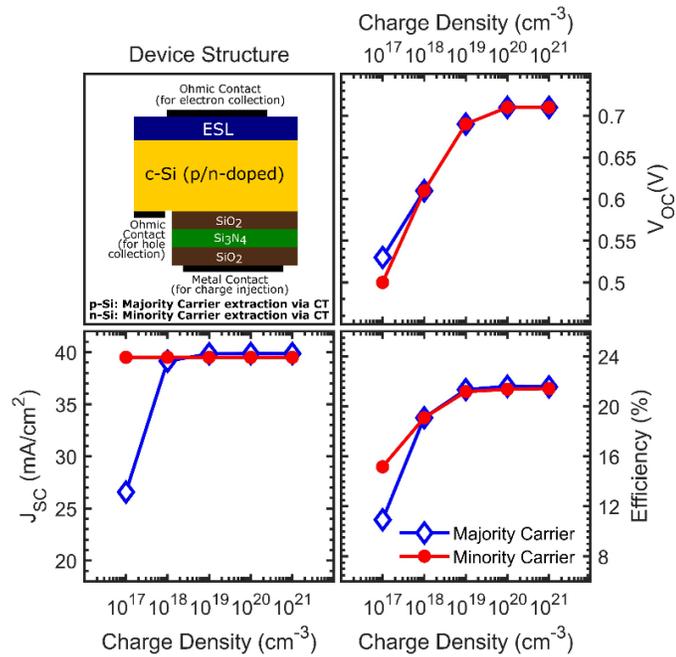

Fig. 6: Effect of charge density on device performance for a solar cell with Si/ESL heterojunction as an electron selection scheme and charges ONO stack as hole selective scheme. The charges stack is explored as a hole selection mechanism for a p-Si substrate, when it functions as a majority carrier selection scheme and for a n-substrate Si where it functions as a minority carrier selection scheme.

We will now look at a solar cell structure (fig. 7) which uses a doped homojunction as an electron extracting scheme and a charge-based hole extraction. The results of fig. 7 show that contrary to what was observed for the case of a Si/TMO based electron extraction, when the front electron selective structure is a homojunction (Fig. 7), the performance strongly depends on whether the charge trap is functioning as a minority or a majority carrier extraction scheme. The charge-based hole

### C. Effect of Area Fraction

The area fraction denotes the ratio of the contacted area to the total rear area of the device. The results of fig. 8 indicate that even when an Ohmic contact is used, a reduction in contacted area can significantly improve device performance. Interestingly, while a low contacted area fraction, though partially compensating for lack of hole selectivity is incapable of providing high efficiencies. The results also show that it is only after the introduction of charge in the dielectric stack the small area fraction can deliver competitive efficiencies. Finally, we observe that for the scheme to be effective, the area fraction must be at most $10^{-4}$. It is noted that an area fraction corresponding to $10^{-5}$ would correspond to one contact of 300 nm X 300 nm for every 100 μm X 100 μm region. Contacts with such dimensions and spacing are routinely defined via photolithography[29] and can even be achieved via laser processing. Hence, the proposed area fraction requirements do not impede the realization of the proposed solar cell.

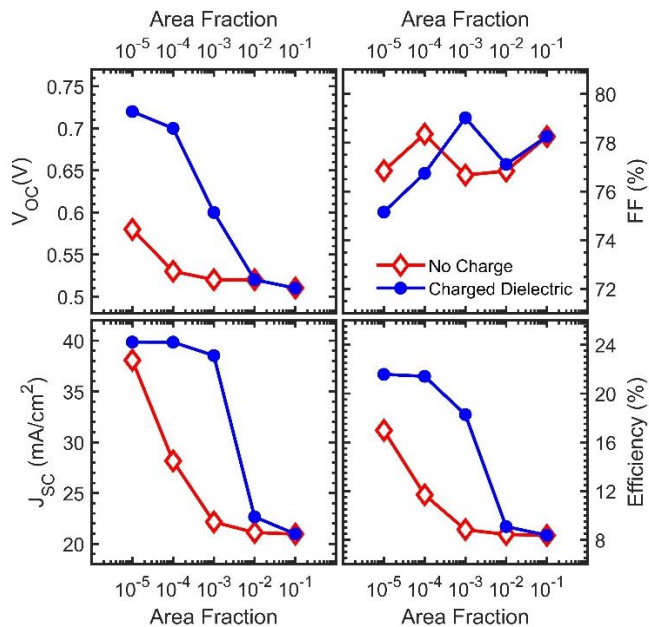

Fig. 8: Effect of area fraction of contacted area on solar cell performance parameters for charge trap with and without negative charge. The negative charge density is kept constant for charged devices at $10^{19}/cm^3$

We note that there have been very limited efforts [30] on having performance recovery methods built in the Si solar cells device. In this context, Chatterji et. al[31] conceptualized a solar cell device with a three terminal physical structure similar to the scheme introduced in this work, which is equipped with performance loss compensation mechanism. The scheme by Chatterji et. al uses the third contact over the oxide is used for applying a continuous potential bias that suppresses recombination and aids charge selectivity by forming an inversion layer. The similarity is restricted to the physical structure only since the working mechanism of the scheme discussed herein is fundamentally different. The current scheme does not require a continuous application of potential to achieve selectivity as the selectivity gets programmed into the device once charge injection into the nitride layer is complete. Furthermore, the recovery mechanism of the scheme also requires constant application of potential, whereas the current scheme only requires periodic charge injection at worst.

We note another study that rigorously explores selectivity based on process introduced interface fixed charge density by Yu et. al[4]. This scheme is also significantly different from the scheme proposed here on two accounts. The first being that this study uses bulk charge for inversion as compared to interface charge discussed by Yu et. al. The second is that the fixed charge in the structure proposed in this work is electrically injected via an additional, third contact and is hence does not have a primary dependence on process parameters.

## IV. CONCLUSIONS

In summary, here we proposed a solar cell structure that uses a negatively charged ONO charge trap that is used in flash memory for achieving hole selectivity. We numerically evaluated the charge trap for its capability to (i) compensate for performance loss arising from deficiency in hole selectivity and passivation, (ii) provide high performance even with inferior passivation of Si by immediate $SiO_2$ and (iii) function as an independent hole selection scheme. The results unequivocally show that the charge trap is capable of all the above and is hence a viable structure for efficient hole collection. Remarkably, it is observed that the scheme can deliver high efficiencies with an SRV as high as 500 cm/s at Si/HSL and Si/$SiO_2$ interface. Finally, the loss compensation capabilities of the structure coupled with the re-programmability of the compensation method make it a highly valuable from the perspective of recovery of on-field performance degradation.

Appendix A: Material Parameters

| Parameter | c-Si (Bulk) | c-Si (Front Doped) | ESL | HSL |
|---|---|---|---|---|
| Bandgap (eV) | 1.12 | 1.12 | 3.5 | 1.12-1.62 |
| Electron Affinity (eV)[32],[33] | 4.05 | 4.05 | 4.02 | 4.05-3.45 |
| $N_C$ ($cm^{-3}$) | 3.23 × $10^{19}$ | 3.23 × $10^{19}$ | 2.5 × $10^{19}$ | 2.5 × $10^{19}$ |
| $N_V$ ($cm^{-3}$) | 1.83 × $10^{19}$ | 1.83 × $10^{19}$ | 2.5 × $10^{19}$ | 2.5 × $10^{19}$ |
| Dielectric Constant | 11.9 | 11.9 | 60 | 60 |
| Tau (SRH) (s) | $10^{-3}$ | $10^{-3}$ | $10^{-6}$ | $10^{-6}$ |
| Radiative Recombination Coefficient ($cm^3 s^{-1}$) | 1.1 × $10^{-14}$ | 1.1 × $10^{-14}$ | -- | -- |
| Auger Recombination Coefficient ($cm^6 s^{-1}$) n,p | 1.0 × $10^{-31}$ 0.79 × $10^{-31}$ | 1.0 × $10^{-31}$ 0.79 × $10^{-31}$ | -- | -- |
| Doping ($cm^{-3}$) n/p | $10^{15}$ n/p | $10^{20}$ n | -- | -- |

Appendix B: Note on physical device realization

The proposed hole selectivity scheme can be implemented by a slight modification of the any partial contact solar cell fabrication process. The front side for both the structures shown in fig. 1(a) and (b) would be standard thin film deposition and dopant diffusion respectively. The dimensions of the spacer between the charge trap and the contact can be dictated by lithography capabilities if the overall area fraction of the contacted region remains in the range discussed in section IIIc.

This would be followed by deposition of the dielectric ONO stack via PECVD (Plasma Enhanced Chemical Vapor Deposition). The stack would be patterned via photolithography. The resist developed for this step would be used as mask for hole contact metal deposition. Post lift-off, a patterned ONO and hole contact would be achieved. Finally, the charging contact would be deposited by second photolithography step where the developed resist is used as mask at the patterned contact is obtained after lift-off. The added processing of 2 levels photolithography promises better reliability and potential elimination of a high temperature doping step for n-type PERC and wider range of transition metal oxides to select from for passivating hole selective layers. Hence, the proposed structure has the potential to be a low-cost, low-temperature-processed, highly efficient, and stable solar cell with less stringent criteria to choose materials to be used for selectivity and passivation.